\documentclass[11pt,twoside]{article}
\usepackage{asp2006}
\usepackage{epsf}
\usepackage{psfig}
\usepackage{lscape}
\usepackage{graphicx}
\begin{document}
\title{Variability-selected low luminosity AGNs in the SA57 and in the CDFS}   
\author{F. Vagnetti}   
\affil{Universit\`a di Roma Tor Vergata, Dipartimento di Fisica, via della Ricerca Scientifica 1, I-00133 Roma, Italy}    
\author{K. Boutsia}   
\affil{INAF-Osservatorio Astronomico di Roma, via di Frascati 33, I-00040 Monte Porzio Catone, Italy} 
\author{D. Trevese}   
\affil{Universit\`a di Roma La Sapienza, Dipartimento di Fisica, P.le A. Moro 2, I-00185 Roma, Italy}    

\begin{abstract} 
Low Luminosity Active Galactic Nuclei (LLAGNs) are contaminated by the light of their host galaxies, thus they cannot be detected by the usual colour techniques.  For this reason their evolution in cosmic time is poorly known. Variability is a property shared by virtually all active galactic nuclei, and it  was adopted as a criterion to select them using multi epoch surveys. Here we report on two variability surveys in different sky areas, the Selected Area 57 and the Chandra Deep Field South. 
\end{abstract}

\section{Introduction}
The knowledge of the luminosity function of Active Galactic Nuclei (AGNs) is based on the construction of statistically significant samples, selected via different techniques derived by the AGN properties, such as non-stellar colour, broad emission lines, or variability. The comparison between samples selected using different techniques enables to evaluate the relevant selection effects and to derive the intrinsic cosmological evolution of the AGN population. 
Colour-selection is limited to point-like objects, i.e. bright  active nuclei outshining the host galaxy, which would  otherwise produce non-stellar colours. 
The COMBO-17 survey \citep{wolf03} provided a ``low resolution spectrum''  enabling to drop the ``point-like" condition, for selecting AGNs on the basis of their  Spectral Energy Distribution (SED) alone. Even in this case, however, the selection is limited to nuclei brighter than $L_B\simeq 10^{44}$ erg s$^{-1}$, since otherwise the SED is dominated by the host galaxy light, which prevents  the nuclear spectrum from being recognised.
X-ray surveys, although efficient in detecting standard AGNs, can miss relatively X-ray weak objects. These could amount to $\sim$ 2\% of the quasars at the level X/O=0.1 \citep{gibs08}, while LLAGNs, intrinsically weak in both optical and X-ray bands, can be hardly detected and/or recognized, also due to dilution by host galaxies. Indeed, it is often difficult to discriminate among different types of low luminosity sources, which include Seyfert galaxies, low ionization narrow emission regions (LINERs), star forming galaxies, and normal galaxies. Also for these reasons the luminosity function of LLAGNs is still poorly known.
On the other hand, variability was adopted as a tool for selecting AGNs in many studies \citep[e.g.][]{berg73,t89,cris90,ver95,geh03}. An important aspect of variability as an AGN search technique is that it can be applied to extended objects. These include LLAGNs that cannot be detected by the colour selection since their SED is contaminated by the light of the host galaxy.  In this case, variability selection becomes easier since nuclear variability tends to increase as nuclear luminosity decreases  \citep[e.g.][]{t94}. Variability-selection was first applied to extended objects by \citet{btk98} in the SA 57 (see Section 2).

\section{Selected Area 57}

\begin{figure}[h]
\centerline{\includegraphics[height=7cm]{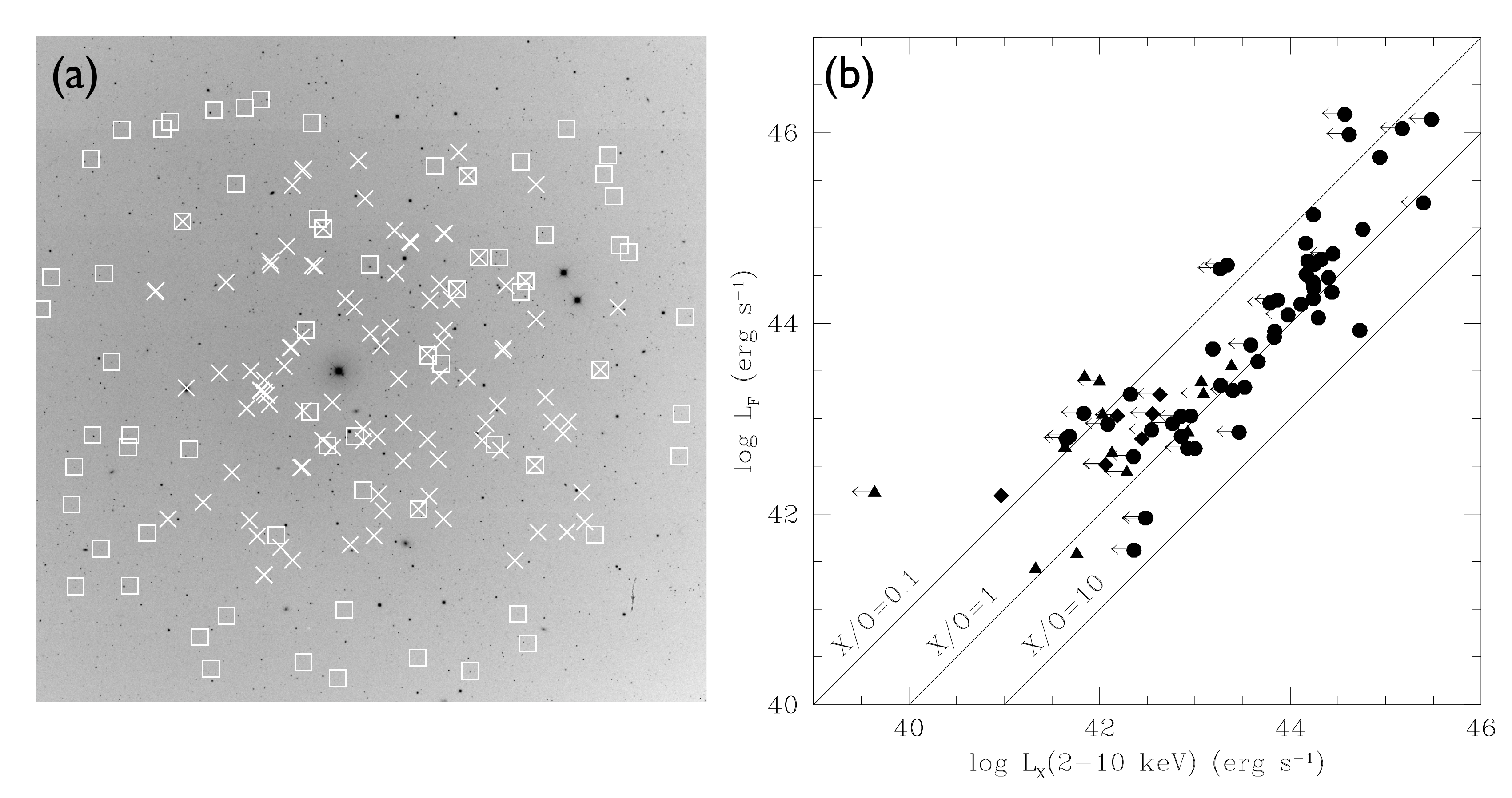}}
%\plotone{vagnetti_fig1.pdf}
\caption{(a) Field of the SA57 with the positions of the variability-selected AGN candidates, marked by squares, and of the X-ray sources, marked by crosses. (b) $F$-band optical luminosity vs 2-10 keV X-ray luminosity for the sources in SA57. Circles: broad-line AGNs; diamonds: narrow emission-line galaxies; triangles: normal galaxies.}
\end{figure}

\noindent
The Selected Area 57 is a well studied area in the optical since the '70s with a number of search techniques,  including non-stellar  colour, absence of  proper motion and variability \citep{kkc86,koo88,t89,t94,btk98}. A field of $\sim  35$ arcmin in diameter has been repeatedly observed  since 1975 in the $U$, $B_J$, $F$, $N$ bands. 
A sample of "variable galaxies", i.e. variable objects with extended images, was created from the same plates of SA 57 \citep{btk98}.
A field of $\sim 0.2$ deg$^2$ in the SA 57 was observed for 67 ks with XMM-Newton (Trevese et al 2007). A catalogue of 140 X-ray sources was created and 98 of them were identified by cross-correlating the X-ray sample with the photometric catalogue of  8146 objects by the Kitt Peak National Observatory \citep{kron80,koo86}. The X-ray catalogue includes 9 sources previously selected on the basis of their variability, while 21 other variability-selected sources are not detected in X-rays. Follow-up spectroscopy was performed at 4.2m William Herschel Telescope and 3.5m Telescopio Nazionale Galileo at La Palma \citep{trev08a}. Figure 1a shows the optical field with the positions of the variability-selected and X-ray-selected sources. 
Figure 1b compares $F$-band optical luminosities and 2-10 keV X-ray luminosities for variability- and/or X-ray-selected sources which have a spectroscopic redshift. Broad-line AGNs are represented as circles, while diamonds and triangles indicate narrow emission-line galaxies and normal galaxies, respectively. AGNs are usually found above X-ray/optical ratio X/O=0.1, but a few of them are also found at lower X/O values. Deeper spectroscopy is needed to populate this diagram at fainter fluxes, possibly finding more X-ray weak sources.

\section{Chandra Deep Field South}

\begin{figure}[h]
\centerline{\includegraphics[height=7cm]{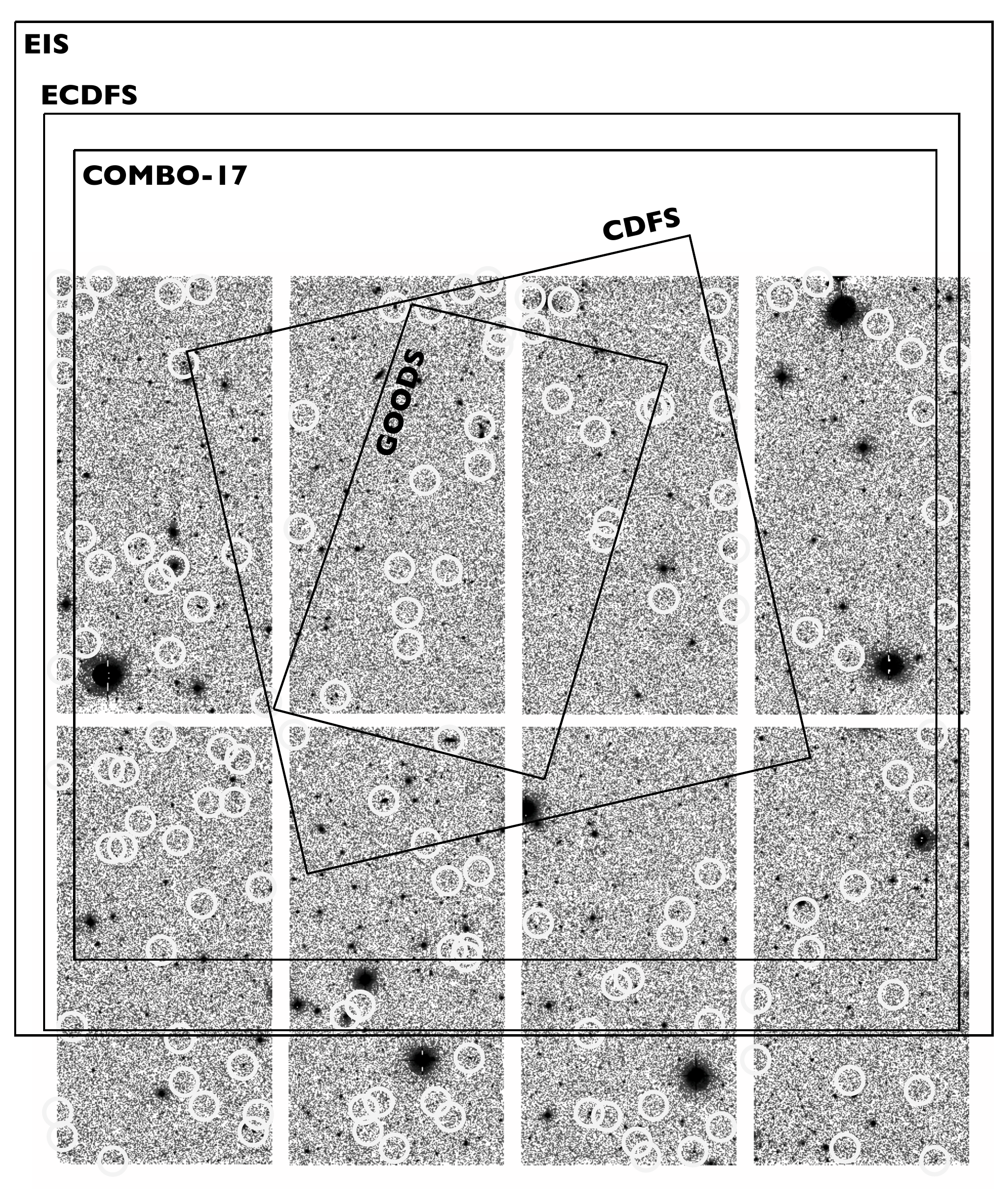}}
\caption{The AXAF field of the STRESS survey, with the positions of the variability-selected AGN candidates, marked by circles. The borders of the overlapping fields of the EIS, COMBO-17, ECDFS, CDFS, GOODS surveys are also shown.}
\end{figure}

\noindent
Variable AGN candidates can be selected in synergic analyses of surveys devoted to the search of supernovae. The Southern inTermediate Redshift ESO Supernova Search (STRESS) \citep{bott08} includes 21 fields (16 with multi-band information), each $\sim 0.3$ deg$^2$, monitored for about 2 years with the ESO/MPI 2.2m telescope. The AXAF field of the STRESS project was chosen to search for AGN candidates though variability\citep{trev08b}. The field is centred at $\alpha$=03:32:23.7, $\delta$=-27:55:52 (J2000) and overlaps with the Chandra Deep Field South (CDFS) \citep{giac02} and other well known surveys, such as the COMBO-17 survey \citep{wolf03}, the ESO Imaging Survey (EIS) \citep{arno01}, the Extended-CDFS survey (ECDFS) \citep{lehm05}, the GOODS survey \citep{giav04}. The images obtained by the STRESS survey in the V band at 8 epochs were analyzed, producing a catalogue of 7267 objects observed at least at 5 epochs. A r.m.s. variability measure $\sigma$ was computed for each object, selecting as variable those objects having $\sigma$ greater than a given threshold. Details of the procedure, as well as the catalogue of the 132 variable objects, are reported in \citet{trev08b}. The STRESS/AXAF field is shown in Figure 2, with the positions of the 132 variability-selected objects, and with the borders of the overlapping fields from other surveys. Follow-up spectroscopy was then performed with the ESO/NTT at La Silla for objects brighter than $V\sim 21.3$ \citep{bout09}.
Figure 3a shows the $R$-band optical luminosities and the 2-8 keV X-ray luminosities for the sources having a spectroscopic redshift. Figure 3b shows the relation between the r.m.s. variability $\sigma$ and the 2-8 keV X-ray luminosity for the same sources. Variable objects are represented as circles (broad-line AGNs), diamonds (narrow emission-line galaxies), and squares (normal galaxies); small dots represent non-variable objects. Concerning low-luminosity sources (narrow emission-line galaxies and normal galaxies), it is possible to separate two groups: one with higher X/O ratio and variability (indicated in the figure by continuous contours), and another with lower X/O ratio and variability (indicated by dashed contours). Objects of the first group have also $U-B$ and $B-V$ colours consistent with normal quasars, while the colours of the second group are dominated by the host galaxies \citep{bout09}. All the properties of the objects belonging to the second group are consistent with LLAGNs diluted by host galaxy light.

\begin{figure}[h]
\centerline{\includegraphics[height=7cm]{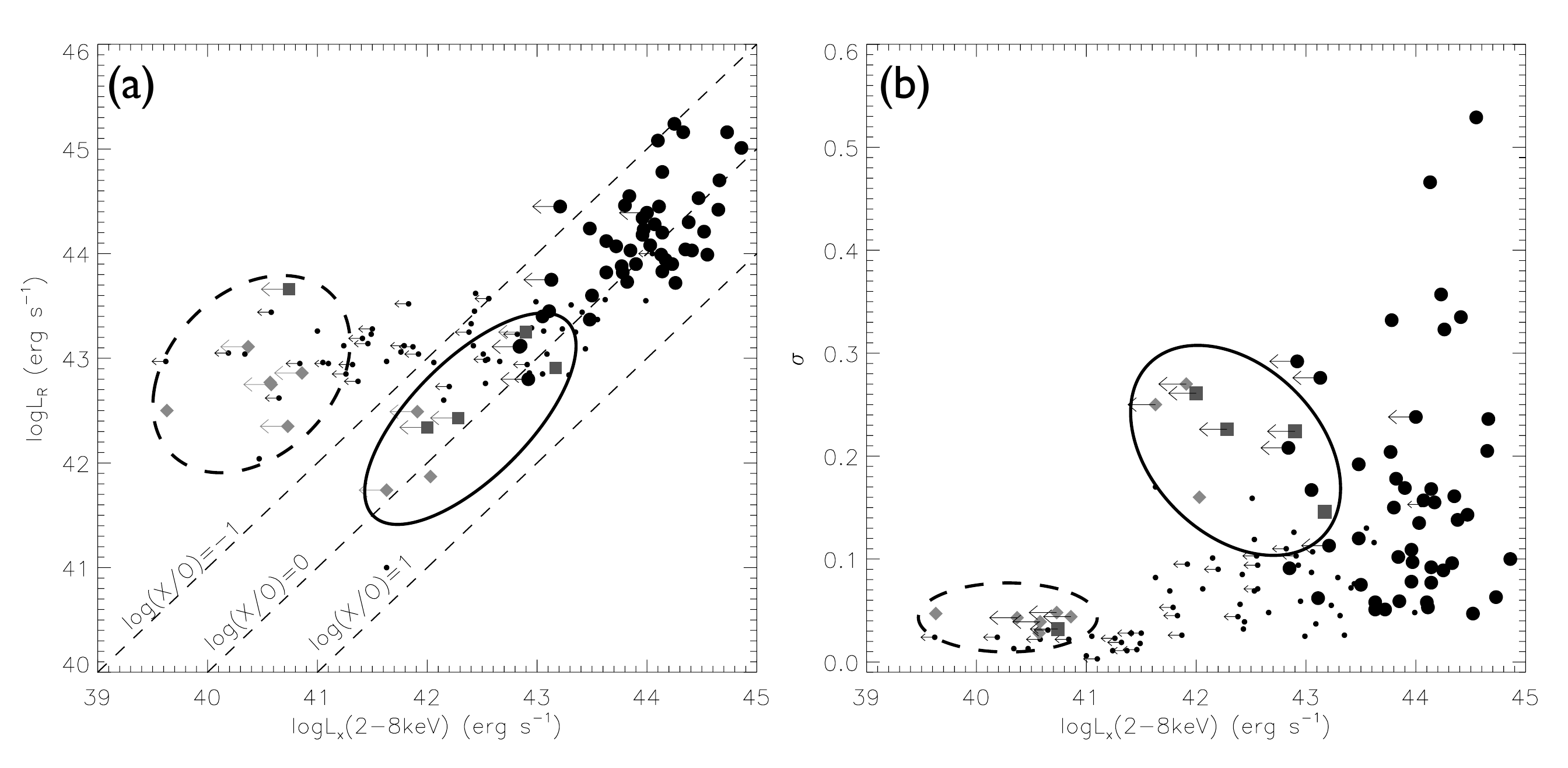}}
\caption{(a) $R$-band optical luminosity vs 2-8 keV X-ray luminosity for the sources with spectroscopic redshift. (b) r.m.s. variability $\sigma$ vs 2-8 keV X-ray luminosity for the same sources. Circles, diamonds, and squares represent variability-selected broad-line AGNs, narrow emission-line galaxies, and normal galaxies, respectively; small dots represent non-variable objects. Continuous and dashed contours indicate the groups of low-luminosity objects discussed in the text.}
\end{figure}

\section{Conclusions}
\noindent
\begin{itemize}
\item[-] LLAGNs are important to understand the evolution of the AGN luminosity function at faint end;
\item[-] variability is an efficient and reliable method to detect LLAGNs lost by colour techniques and by X-ray surveys;
\item[-] it is crucial to discriminate among the different classes of low luminosity objects;
\item[-] SA 57 is the first field where extended variable objects have been selected;
\item[-] deeper spectroscopy of faint ($B<23$) extended variable candidates is needed;
\item[-] synergy in variability searches (SNe, AGNs) can provide statistical samples of  
low luminosity AGNs;
\item[-] 132 variability-selected AGN candidates were found in the STRESS/AXAF field;
\item[-] some variable objects (7/132) with low X/O ratio appear consistent with LLAGNs diluted by the host galaxy light.
\end{itemize}


\begin{thebibliography}{}
\bibitem[Arnouts et al.(2001)]{arno01}Arnouts, S., Vandame, B., Benoist, C., et al. 2001, A\&A 379, 740
\bibitem[Bershady et al.(1998)]{btk98}Bershady, M. A., Trevese, D., \& Kron, R. G. 1998, ApJ 496, 103 
\bibitem[Botticella et al.(2008)]{bott08}Botticella, M. T., Riello, M., Cappellaro, E. et al. 2008,A\&A 479, 49
\bibitem[Boutsia et al.(2009)]{bout09}Boutsia, K., Leibundgut, B., Trevese, D., \& Vagnetti, F. 2009, A\&A, in press, arXiv:0902.0933
\bibitem[Cristiani et al.(1990)]{cris90}Cristiani, S., Vio, R., Andreani, P. 1990, AJ 100, 56
\bibitem[Geha et al.(2003)]{geh03}Geha, M., Alcock, C., Allsman, R. A., et al. 2003, ApJ 125,1
\bibitem[Giacconi et al.(2002)]{giac02}Giacconi, R., Zirm, A., Wang, J. 2002, ApJS 139, 369 
\bibitem[Giavalisco et al.(2004)]{giav04}Giavalisco, M., Ferguson, H.C., Koekemoer, A.M. 2004, ApJ 600, 93
\bibitem[Gibson et al.(2008)]{gibs08}Gibson, R.R., Brandt, W.N., \& Schneider, D.P. 2008 ApJ 685, 773
\bibitem[Koo(1986)]{koo86} Koo, D.C. 1986, ApJ, 311, 651
\bibitem[Koo \& Kron(1988)]{koo88} Koo, D.C., \& Kron, R.G. 1988, ApJ 325, 92 
\bibitem[Koo et al.(1986)]{kkc86}Koo, D. C., Kron, R. G., \& Cudworth, K. M. 1986, PASP 98, 285
\bibitem[Kron(1980)]{kron80}Kron, R. G.	1980, ApJS, 43, 305
\bibitem[Lehmer et al.(2005)]{lehm05}Lehmer, B.D., Brandt, W.N., Alexander, D.M., et al. 2005, ApJS 16, 21
\bibitem[Trevese et al.(1989)]{t89}Trevese, D., Pittella G., Kron R. G.,Koo D. C., \& Bershady M. A. 1989, AJ 98, 108
\bibitem[Trevese et al.(1994)]{t94}Trevese, D., Kron, R. G., Majewski, S. R., Bershady, M. A., Koo, D. C. 1994, ApJ 433, 494
\bibitem[Trevese et al.(2007)]{trev07}Trevese, D., Vagnetti, F., Puccetti, S., Fiore, F., Tomei, M., Bershady, M. A. 2007, A\&A 469, 1211 
\bibitem[Trevese et al.(2008a)]{trev08a}Trevese, D., Vagnetti, F., Zitelli, V., Boutsia, K., \& Stirpe, G.M. 2008a, A\&A 477, 473
\bibitem[Trevese et al.(2008b)]{trev08b}Trevese, D., Boutsia, K., Vagnetti, F., Cappellaro, E., Puccetti, S. 2008b, A\&A 488, 73
\bibitem[van den Berg et al.(1973)]{berg73}van den Berg, S., Herbst, E., \& Pritchet, C. 1973, AJ  78, 375
\bibitem[V\'eron \& Hawkins(1995)]{ver95}V\'eron, P., \& Hawkins, M.R.S. 1995, A\&A 296, 665
\bibitem[Wolf et al.(2003)]{wolf03}Wolf, C., Wisotzki, L., Borch, A.,  et al. 2003, A\&A 408, 499
\end{thebibliography}
\end{document}